\documentclass[
  aps,
  pra,
  twocolumn,
  superscriptaddress,
  floatfix,
  longbibliography
]{revtex4-2}

\usepackage{amsmath,amssymb,amsfonts}
\usepackage{graphicx}
\usepackage{hyperref}
\usepackage{xcolor}
\usepackage{braket}
\usepackage{booktabs}
\usepackage{dcolumn}

\newcommand{\KLD}{D_{\mathrm{KL}}}
\newcommand{\NLL}{\mathcal{L}_{\mathrm{NLL}}}
\newcommand{\VarGrad}{\mathrm{Var}[\partial_k\mathcal{L}]}
\newcommand{\Us}{\hat{U}_{S}}
\newcommand{\Uhaar}{\hat{U}_{\mathrm{Haar}}}

\newcommand{\Ua}{\hat{U}_{a}}

\newcommand{\NA}{N_A}
\newcommand{\NQ}{N}
\newcommand{\sx}{\hat{\sigma}^x}
\newcommand{\sy}{\hat{\sigma}^y}
\newcommand{\sz}{\hat{\sigma}^z}
\newcommand{\Ho}{\hat{H}}

\graphicspath{{./}}

\begin{document}

\title{Quantum Scrambling Born Machine}

\author{Marcin P\l{}odzie\'{n}}
\email{marcin.plodzien@qilimanjaro.tech}
\affiliation{Qilimanjaro Quantum Tech, Carrer de Vene\c{c}uela 74, 08019 Barcelona, Spain}

\date{\today}

\begin{abstract}
We propose a Quantum Scrambling Born Machine (QSBM) that decouples entanglement generation from the trainable degrees of freedom: a fixed, non-trainable scrambling unitary provides multi-qubit entanglement while only single-qubit rotations are optimized.
We consider three entangling unitaries (a Haar random unitary and two physically realizable approximations, a finite-depth brickwork random circuit and analog time evolution under nearest-neighbor spin-chain Hamiltonians), and show that, for the benchmark distributions and system sizes considered, once the entangler produces near-Haar-typical entanglement the model learns the target distribution with weak sensitivity to the scrambler's microscopic origin.
Because the scrambler requires only single-shot calibration, the entire trainable parameter budget is devoted to single-qubit rotations, eliminating the need for trainable multi-qubit entangling gates on near-term hardware.
Finally, promoting the Hamiltonian couplings to trainable parameters casts the generative task as a variational Hamiltonian problem, with performance competitive with representative classical generative models at matched parameter count.
\end{abstract}

\maketitle

\section{Introduction}\label{sec:introduction}

\begin{figure}[t]
    \centering
    \includegraphics[width=\columnwidth]{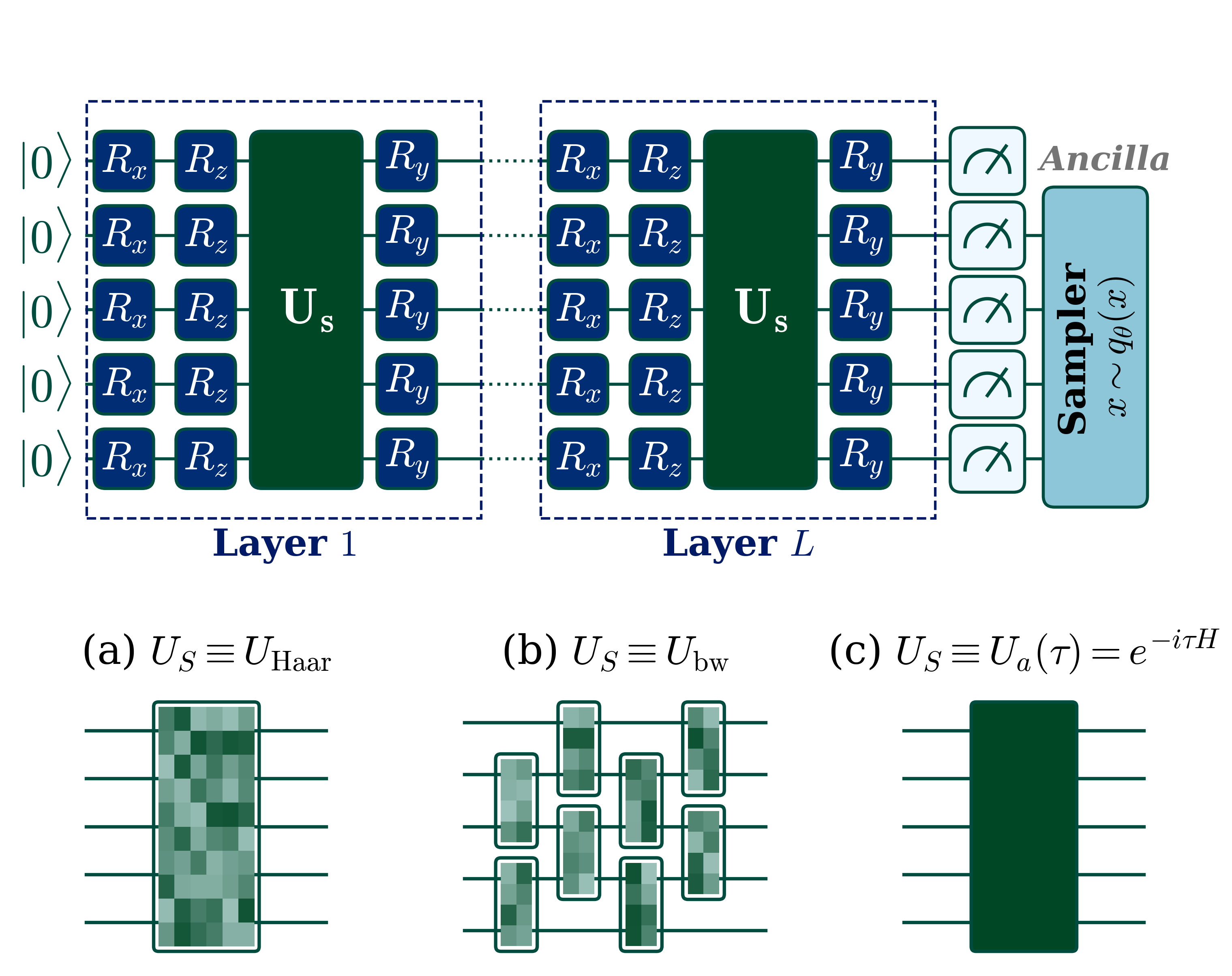}
    \caption{Architecture of the quantum scrambling Born machine.
    An $\NQ$-qubit register, initialized in $\ket{0}^{\otimes \NQ}$, is processed through $L$ repeated layers; each layer applies trainable single-qubit rotations ($\hat{R}_x, \hat{R}_z$ before, $\hat{R}_y$ after) surrounding a fixed scrambling unitary~$\Us$ that generates entanglement.
    After $L$ layers, $\NA$ ancilla qubits are traced out and the remaining $n = \NQ - \NA$ system qubits are measured in the computational basis, producing a probability distribution via the Born rule [Eq.~\eqref{eq:born_rule}].
    Only the rotation angles are optimized; the scrambler is kept fixed.
    Three scrambler types are considered: {(a)}~Haar random unitary, {(b)}~brickwork random quantum circuit (RQC) of depth~$K$, {(c)}~analog Hamiltonian evolution $e^{-i\tau \Ho}$.}
    \label{fig:architecture}
\end{figure}

Quantum machine learning (QML) exploits quantum resources (superposition, entanglement, and the exponential dimensionality of Hilbert space) to enhance learning tasks~\cite{biamonte2017quantum, schuld2021machine, dunjko2020nonreview, dawid2024machine, conti2024quantum, cerezo2021variational, wang2024comprehensive, gujju2024quantum, peralgarcia2024systematic}, with quantum generative models emerging as a promising near-term application~\cite{benedetti2019generative, zoufal2021generative, kasture2023protocols, delgado2025quantum, chen2025quantum}.
Quantum generative models take inspiration from their classical counterparts, such as quantum generative adversarial networks (qGANs)~\cite{dallaire2018quantum, zoufal2019quantum, niu2022entangling, gao2022enhancing} and quantum diffusion models~\cite{parigi2024quantum}.

A conceptually distinct family are quantum circuit Born machines (QCBMs), in which a parametrized quantum state is measured in the computational basis, with the Born rule naturally defining a normalized probability distribution over exponentially many outcomes~\cite{liu2018differentiable, cheng2018information, coyle2020born}.
Born machines can provably achieve an exponential representational advantage over classical generative models~\cite{gao2018quantum}, and subsequent work has benchmarked QCBMs on financial and synthetic datasets, including their generalization properties~\cite{rudolph2024trainability, gili2024generalization, riofrio2024performance, gili2024generalize}.

Quantum Born machines, like other variational quantum algorithms, face a tradeoff between expressivity and trainability of the parametric ansatz~\cite{cerezo2021variational}.
Highly expressive circuits, such as deep random parametric architectures, can in principle represent any target distribution; however, they suffer from barren plateaus, where gradients of the loss function vanish exponentially with system size, rendering optimization infeasible~\cite{mcclean2018barren, cerezo2021cost, holmes2022connecting}. By the Lie algebraic theory of barren plateaus~\cite{ragone2024liealgebraic}, this bottleneck is set by the dynamical Lie algebra (DLA) of the circuit; the QSBM does not evade it, but removes the dominant hardware cost---trainable multi-qubit entangling gates---by fixing the entangler.
Shallow or overly structured circuits may lack the entangling power to represent complex target distributions faithfully~\cite{sim2019expressibility, du2022efficient}.

In this work we propose a Quantum Scrambling Born Machine (QSBM), which prioritizes hardware and parameter efficiency for near-term quantum processors by decoupling entanglement generation from the trainable degrees of freedom. We treat quantum scrambling---the rapid spread of initially localized quantum information across all degrees of freedom of a many-body system~\cite{hayden2007black, sekino2008fast, swingle2018unscrambling, hosur2016chaos, lomonaco2025operational, plodzien2025scrambling}---as an entangling resource for quantum generative modeling: once the fixed unitary produces Haar-typical bipartite entanglement,  diagnosed by the half-chain von Neumann entropy approaching the Page value \cite{page1993average}, representation learning is controlled entirely through single-qubit rotations, see Fig.~\ref{fig:architecture}.
The entangling layer is provided by either an ideal Haar random unitary or its physical approximations---finite-depth brickwork random circuits on digital platforms and finite-time Hamiltonian evolution on analog quantum simulators.
Because the scrambler is a fixed, non-trainable unitary reused in every layer, it needs to be calibrated or compiled only once; the entire trainable parameter budget is devoted to single-qubit rotations, making the architecture particularly suited to near-term hardware where multi-qubit gate recalibration is the dominant overhead.

We monitor the generative performance from under-scrambled to fully scrambled regimes and find that, for the qubit counts ($\NQ \le 10$) and target families studied here, the converged model quality is controlled by how closely the entangler's bipartite entanglement approaches the Haar-typical value. For a fixed qubit budget, tracing out ancilla qubits increases expressivity by exploring mixed-state distributions. Extending the architecture to trainable Hamiltonian couplings casts the generative task as a variational Hamiltonian problem, complementing the inverse Hamiltonian learning problem~\cite{huang2023heisenberg, yu2023robust, dutkiewicz2024advantage, gu2024practical, kokail2021entanglement, heightman2025hamiltonian, heightman2026lindbladian, olsacher2025hamiltonian, franceschetto2025zeno, baran2025heisenberg, guo2025hamiltonian, kraft2025bounded}. A comparison with representative classical generative models at matched parameter count indicates that the QSBM is competitive in the small trainable-parameter regime.

Standard QCBMs include trainable entangling layers~\cite{liu2018differentiable, cheng2018information, coyle2020born}; in the QSBM, by contrast, the entangler is fixed and all optimization is confined to single-qubit rotations.

The remainder of the paper is organized as follows.
Section~\ref{sec:model} introduces the model architecture, the role of scrambling and Haar typicality.
Section~\ref{sec:training_protocol} describes the training protocol, loss function, and target distributions.
Section~\ref{sec:results} presents the main results for Haar, digital, and analog scramblers.
Section~\ref{sec:2d} extends the architecture to trainable Hamiltonian parameters and benchmarks it against classical generative models.
We conclude in Sec.~\ref{sec:conclusions}.

\section{Model}\label{sec:model}

\subsection{Quantum Born machines}\label{sec:born_machine}

A quantum Born machine is a generative model that encodes a probability distribution in the squared amplitudes of a quantum state~\cite{liu2018differentiable, cheng2018information}.
Given $\NQ$ qubits, we prepare a parameterized state by applying $L$ identical structural layers to the reference state $\ket{0}^{\otimes \NQ}$.
Each layer~$\ell = 1,\ldots,L$ consists of trainable single-qubit pre-rotations $\hat{R}_x(\theta^{(\ell)}_{i,x})\, \hat{R}_z(\theta^{(\ell)}_{i,z})$ applied to every qubit~$i$, followed by a fixed scrambling unitary~$\Us$, and concluded by a trainable post-rotation $\hat{R}_y(\theta^{(\ell)}_{i,y})$ on every qubit.
The full circuit unitary reads
\begin{equation}\label{eq:ansatz}
    \hat{U}(\boldsymbol{\theta}) = \prod_{\ell=1}^{L} \hat{R}^{(\ell)}_{y,\mathrm{post}}\, \Us\, \hat{R}^{(\ell)}_{z,\mathrm{pre}}\, \hat{R}^{(\ell)}_{x,\mathrm{pre}},
\end{equation}
where the product is ordered with $\ell = 1$ applied first (rightmost), the rotation operators act independently on each qubit, $\hat{R}^{(\ell)}_{\alpha,\mathrm{pre/post}} = \bigotimes_{i=1}^{\NQ} \hat{R}_{\alpha}(\theta^{(\ell)}_{i,\alpha})$, and each angle $\theta^{(\ell)}_{i,\alpha}$ is an independently trained parameter.
The same scrambling unitary $\Us$ is reused in every layer, acting as a fixed entangling block that propagates multi-qubit entanglement.
Reuse also reflects the hardware-relevant scenario in which the entangling operation is calibrated or compiled once and held fixed, so that only single-qubit control pulses are adjusted during training.
What distinguishes successive layers are their independently trained rotation angles.
The pre-rotations $\hat{R}_x$, $\hat{R}_z$, then fixed scrambler $\Us$, then post-rotation $\hat{R}_y$ serves as a minimal universal local frame; alternative orderings yield qualitatively similar performance.
The resulting state $\ket{\psi(\boldsymbol{\theta})} = \hat{U}(\boldsymbol{\theta}) \ket{0}^{\otimes \NQ}$ is measured in the computational basis.
For $\NA = 0$ (no ancillas), every qubit is measured and outcome $x \in \{0,1\}^{\NQ}$ occurs with probability
\begin{equation}\label{eq:born_rule}
    q_{\boldsymbol{\theta}}(x) = \big|\braket{x\,|\,\psi(\boldsymbol{\theta})}\big|^2.
\end{equation}
The total number of trainable parameters is $3 L \NQ$ (three rotation angles per qubit per layer), scaling linearly in both $L$ and $\NQ$.
When $\NA$ ancilla qubits are included, they participate in the scrambling but are traced out before measurement. Defining the reduced density matrix $\hat\rho_{\mathrm{sys}} = \mathrm{Tr}_{A}\!\big[\ket{\psi(\boldsymbol{\theta})}\!\bra{\psi(\boldsymbol{\theta})}\big]$, the output probability is
\begin{equation}\label{eq:ancilla_trace}
    q_{\boldsymbol{\theta}}(x) = \langle x | \hat \rho_{\mathrm{sys}} | x \rangle = \sum_{a \in \{0,1\}^{\NA}} |\braket{x, a\,|\,\psi(\boldsymbol{\theta})}|^2,
\end{equation}
where $x \in \{0,1\}^n$ labels computational-basis states of the measured register of size $n = \NQ - \NA$.
Since $\hat\rho_{\mathrm{sys}}$ has rank at most $2^{\NA}$, its spectral decomposition expresses the output distribution as a convex combination of up to $2^{\NA}$ pure-state Born distributions, enlarging the representable class beyond a single pure state.

\subsection{Quantum scrambling and Haar typicality}\label{sec:haar_theory}

Entanglement and many-body correlations are essential for the expressivity of quantum generative models, since a product-state circuit can only produce factorized output distributions.
The ideal scrambling unitary is a Haar-random unitary $\Uhaar$, drawn from the unique unitarily invariant measure on $U(2^{\NQ})$~\cite{mele2024introduction, plodzien2025scrambling}, which maps an initial product state $\ket{0}^{\otimes \NQ}$ to a state whose quantum information is delocalized across all degrees of freedom~\cite{hayden2007black}.
A key property of such states is typicality~\cite{bengtsson2017geometry, hayden2006aspects}, i.e. almost every Haar-random state is nearly maximally entangled across any bipartition, with the average bipartite entropy given by the Page formula~\cite{page1993average}, $\langle S_A \rangle_{\mathrm{Haar}} = \sum_{k=d_B+1}^{d_A d_B} k^{-1} - (d_A - 1)/(2d_B) \approx \ln d_A$ (where $d_A d_B = 2^{\NQ}$, $d_B \ge d_A$), and fluctuations around this value exponentially suppressed in $\NQ$.
When $\NA$ ancilla qubits are traced out from a Haar-scrambled state, the reduced density matrix $\hat\rho_{\mathrm{sys}}$ follows the fixed-trace Wishart distribution~\cite{zyczkowski2001induced} with generically non-degenerate eigenvalues.

Implementing a true Haar-random unitary requires a gate count that grows with $\NQ$~\cite{mezzadri2007generate}, making it impractical beyond small system sizes.
Two physically realizable strategies approximate Haar-level scrambling.
(i)~Random quantum circuits of depth $K$ in a brickwork layout converge to approximate unitary $t$-designs---ensembles whose first $t$ moments match those of the Haar measure.
Already at $t = 2$, coarse entanglement properties such as subsystem purities and R\'{e}nyi-2 entropies become indistinguishable from the Haar ensemble~\cite{harrow2009random, brandao2016local, haah2024efficient, harrow2023approximate, fefferman2024anti, foxman2025random}.
(ii)~Analog time evolution under a many-body Hamiltonian produces effective scrambling at sufficiently long times, with the entanglement entropy approaching the Page value~\cite{emerson2003pseudo, maldacena2016bound, plodzien2025scrambling}.

Both approximations are expected to produce Haar-typical entanglement at sufficient depth or evolution time, with subsystem entropies converging to the Page value.
In practice, we benchmark each scrambler against the generative performance of a Haar-random unitary $\Us = \Uhaar$.

\section{Training and evaluation protocol}\label{sec:training_protocol}

We train the Born machine by minimizing the negative log-likelihood (NLL),
\begin{equation}\label{eq:nll}
    \NLL(\boldsymbol{\theta}) = -\sum_{x} p(x) \ln q_{\boldsymbol{\theta}}(x),
\end{equation}
where $p(x)$ is the target distribution.  The NLL is related to the Kullback--Leibler divergence (KLD) by $\KLD(p \| q_{\boldsymbol{\theta}}) = \NLL(\boldsymbol{\theta}) - H(p)$, where $H(p)$ is the Shannon entropy of the target~\cite{cover2006elements}.
Since $H(p)$ is constant, minimizing NLL is equivalent to minimizing $\KLD$, our primary performance metric.

For the one-dimensional benchmarks (Secs.~\ref{sec:haar}--\ref{sec:analog}), we employ a multimodal distribution composed of five Gaussian peaks discretized onto $2^n$ bins (where $n = \NQ - \NA$ is the number of measured system qubits):
\begin{equation}\label{eq:multimodal}
    p(x) \propto \sum_{j=1}^{5} w_j \exp\left[ -\frac{(x - \mu_j)^2}{2\sigma^2} \right],
\end{equation}
where the peak centers are uniformly spaced at $\mu_j = (j - 0.5) \cdot 2^n / 5$ for $j = 1,\ldots,5$, the common width is $\sigma = 2^n / 20$, and the weights $w_j$ are drawn randomly from $U(0.5, 1.5)$ with a fixed seed to ensure reproducibility.
The unequal heights and sharp inter-peak valleys make this a non-trivial benchmark for the model.

For the two-dimensional benchmarks (Sec.~\ref{sec:2d}), the $\NQ$ qubits are split into two registers of $n_x = n_y$ system qubits, each mapped to the interval $[-3, 3]$ and discretized onto a $2^{n_x} \times 2^{n_y}$ grid.
We study two target distributions.
(i)~A bivariate Gaussian with tunable correlation,
\begin{equation}\label{eq:2d_gauss}
    p(x,y) \propto \exp\!\left[-\frac{x^2 - 2\rho\, x y + y^2}{2(1 - \rho^2)}\right],
\end{equation}
with $\rho \in \{0, 0.5, 0.9\}$.
(ii)~A mixture of four isotropic Gaussians,
\begin{equation}\label{eq:2d_mixture}
    p(x,y) \propto \sum_{k=1}^{4} \exp\!\left[-\frac{(x - \mu_{x,k})^2 + (y - \mu_{y,k})^2}{2\sigma^2}\right],
\end{equation}
with centers $(\mu_{x,k}, \mu_{y,k}) \in \{(\pm 1.5, \pm 1.5)\}$ and $\sigma = 0.5$.

\section{Quantum scrambling as a generative resource}\label{sec:results}

Unless otherwise noted, all one-dimensional experiments use $\NQ = 8$ total qubits, are trained for 2000 epochs and averaged over 20 independent realizations (each with a freshly drawn scrambler instance and random initial parameters); error bars denote $\pm 1\sigma$.
Optimization uses Adam~\cite{kingma2015adam} (learning rate $0.01$, gradient norm clipping at $1.0$) with exact autodiff gradients computed from the full statevector probabilities.
The reported KLD is evaluated from empirical distributions obtained via $N_{\mathrm{shots}} = 5000$ Born-rule samples.
This finite shot count sets a practical floor on the measurable KLD: even an ideal Haar-random scrambler cannot achieve exactly zero KLD because sampling fluctuations produce a residual divergence of order $1/N_{\mathrm{shots}}$; the horizontal red dashed lines in the figures mark this Haar-typical floor.

\subsection{Haar random unitary}\label{sec:haar}

We begin with the idealized case $\Us = \Uhaar$. Since the scrambler already provides near-maximal entanglement, any remaining limitation in the learned distribution originates from the expressivity of the rotation layers.

\begin{figure}[t]
    \centering
    \includegraphics[width=\columnwidth]{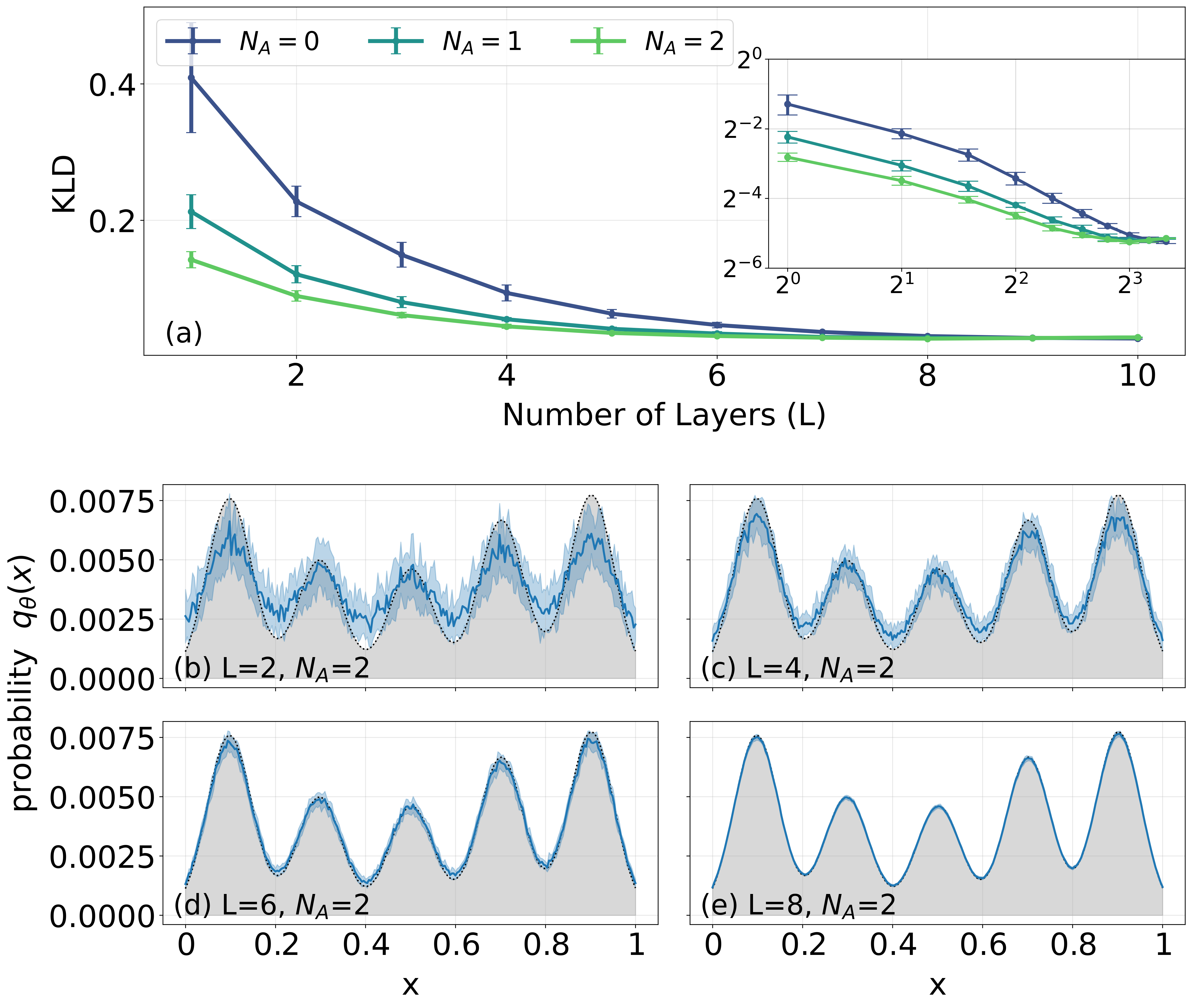}
    \caption{Born machine performance with a Haar-random scrambler ($\NQ = 8$, $N_{\mathrm{shots}} = 5000$).
   {(a)}~Converged KLD vs.\ number of layers~$L$ for $\NA = 0, 1, 2$ ancilla qubits.
    The KLD decreases monotonically with $L$ (inset, log scale), with each additional ancilla qubit lowering the KLD.
   {(b)--(e)}~Target distribution (gray) vs.\ learned Born-machine output (blue) at $\NA = 2$ for $L = 2, 4, 6, 8$, illustrating progressive convergence from a coarse to a near-exact reproduction of the multimodal target.
    Shaded bands: $\pm 1\sigma$ over 20 independent realizations (each with a fresh Haar draw and random initial rotation parameters).}
    \label{fig:haar}
\end{figure}

Figure~\ref{fig:haar}{(a)} shows the converged KLD as a function of the number of layers~$L$, with separate curves for $\NA = 0, 1, 2$ ancilla qubits.
The KLD decreases monotonically with $L$, as the number of trainable rotation gates controls the expressivity at fixed scrambler entanglement. For the pure-state case, $\NA = 0$, the model already learns the target distribution well for $L\ge 6$. However, for shallow circuits, $L\le5$, tracing out ancilla qubits $N_A = 1, 2$ improves the model expressivity.
Panels~{(b)--(e)} illustrate this progression for $\NA = 2$: at $L = 2$ only the coarse envelope of the target is captured, while by $L = 8$ the learned and target distributions agree quantitatively.

Tracing out ancilla qubits simultaneously increases the mixed-state rank of $\hat\rho_{\mathrm{sys}}$ and reduces the output register from $2^n$ to $2^{n-1}$ bins, so the KLD improvement combines a genuine expressivity gain with a lower-dimensional learning task.
At fixed $\NQ$, increasing $\NA$ trades output resolution for mixed-state expressivity; we focus on this hardware-relevant tradeoff.
To isolate the effect of the mixed-state rank from the dimensional reduction, we present a controlled study at fixed output register size $n = 8$ in Appendix~\ref{sec:app_ancilla}, sweeping the number of ancillas $N_A = 0, 1, 2$ (corresponding to total qubit sizes $N_{\mathrm{total}} = 8, 9, 10$). As shown in Fig.~\ref{fig:app_ancilla}, increasing $N_A$ yields a systematic reduction in converged KLD at fixed target resolution, confirming that tracing out scrambled ancilla qubits acts as a genuine expressivity resource.

\subsection{Random quantum circuit of depth $K$}\label{sec:rqc}

Next, we replace the global Haar unitary with a random quantum circuit of variable depth~$K$ arranged in a brickwork topology.
Each brickwork layer applies random two-qubit Haar gates to alternating nearest-neighbor pairs.
Since a circuit of depth $K \propto \NQ$ is expected to converge to an approximate 2-design~\cite{brandao2016local, harrow2009random}, the question is how rapidly the Born machine performance tracks this convergence.

\begin{figure}[t]
    \centering
    \includegraphics[width=\columnwidth]{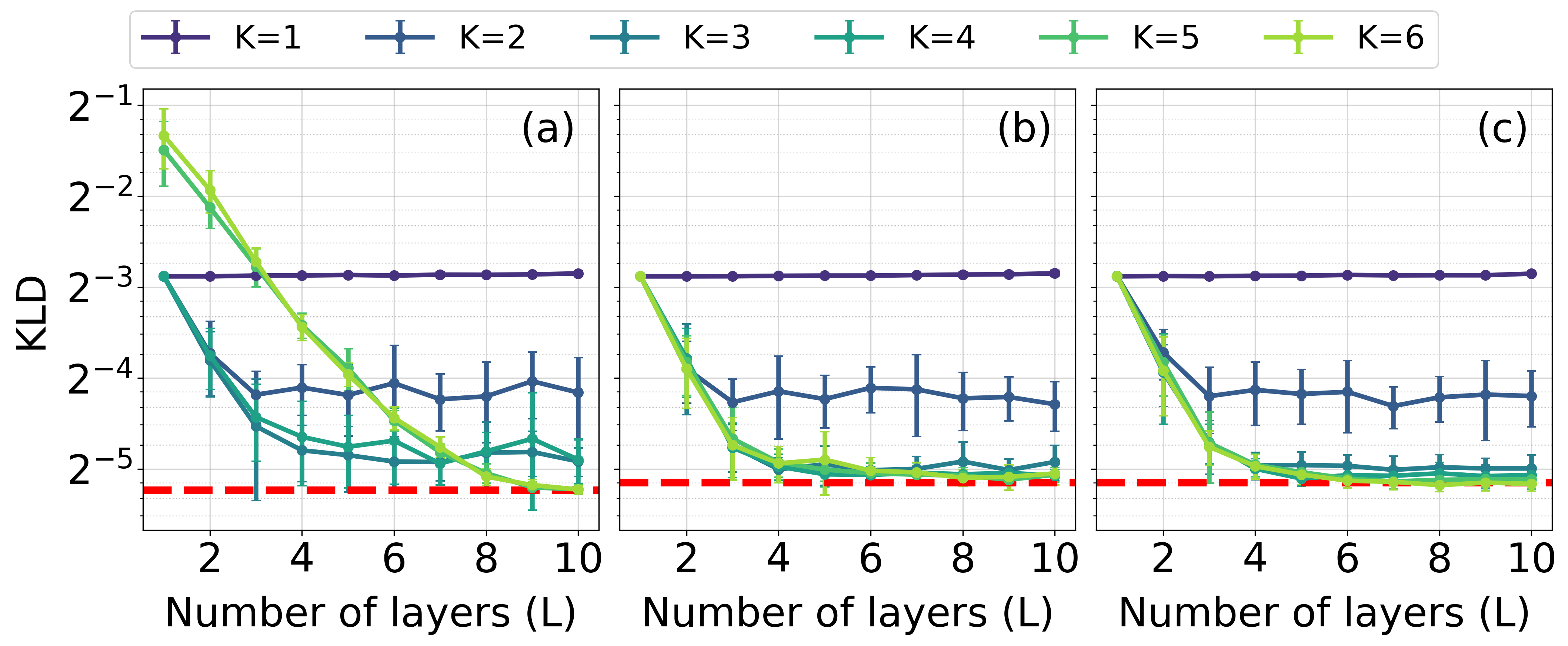}
    \caption{Finite-depth brickwork random circuit scrambler ($\NQ = 8$, $N_{\mathrm{shots}} = 5000$).
    Converged KLD vs.\ number of layers~$L$ for brickwork depths $K = 1$--$6$ at
     {(a)}~$\NA=0$,
     {(b)}~$\NA=1$,
     {(c)}~$\NA=2$.
    At $K = 1$ the circuit generates only nearest-neighbor entanglement and the KLD saturates well above the Haar-scrambler level (red dashed); by $K \ge 3 \approx \NQ/2$ every curve collapses onto the Haar level, consistent with the theoretical expectation that brickwork circuits of depth $O(\NQ)$ converge to approximate 2-designs and reproduce the entanglement properties of a full Haar-random unitary~\cite{brandao2016local}.
    Ancilla qubits lower the KLD.
    Mean $\pm 1\sigma$ over 20 random circuit instances.}
    \label{fig:rqc}
\end{figure}

Fig.~\ref{fig:rqc} presents the converged KLD as a function of layers $L$, for given brickwork circuit depth $K$, with $\NA = 0, 1, 2$ traced-out ancilla qubits (panels {(a--c)}). At $K = 1$ only nearest-neighbor entanglement is generated, insufficient to scramble the full register, and the KLD saturates regardless of $L$.
Already at $K = 2$ the improvement is substantial, and for $K \ge 5$ all curves collapse onto the Haar-scrambler level.
Consistently, the mixed-state expressivity after tracing $N_A$ ancilla qubits improves the model.

\subsection{Analog Hamiltonian scrambler}\label{sec:analog}

In this section we consider the entangling layer as a continuous Hamiltonian evolution, $\Us = \Ua(\tau) = e^{-i\tau \Ho}$, providing an experimentally relevant scrambling resource on analog quantum simulators, where native many-body dynamics serve as the entangling resource, bypassing compiled gate sequences.

We consider the general nearest-neighbor Hamiltonian
\begin{equation}\label{eq:hamiltonian}
    \Ho = -\sum_{\langle i,j \rangle} \left( J_{xx}\, \sx_i \sx_j + J_{yy}\, \sy_i \sy_j + J_{zz}\, \sz_i \sz_j \right) - \sum_i h_x\, \sx_i,
\end{equation}
where $\sx_i, \sy_i, \sz_i$ are the Pauli operators acting on site $i$, the sums $\langle i,j \rangle$ run over nearest neighbors under open boundary conditions, and the parameters $J_{\alpha\alpha}$ and $h_x$ define the interaction and field strengths.

We consider two models: the transverse-field Ising model (TFIM), with $J_{zz} = h_x = 1$,
\begin{equation}\label{eq:H_tfim}
 \Ho_{\mathrm{TFIM}} = -\sum_i \sz_i \sz_{i+1} - \sum_i \sx_i,
\end{equation}
and the XX model with transverse field, $J_{xx} = J_{yy} = h_x = 1$,
\begin{equation}\label{eq:H_xx}
 \Ho_{\mathrm{XX}} = -\sum_i (\sx_i \sx_{i+1} + \sy_i \sy_{i+1}) - h_x\sum_i \sx_i.
\end{equation}
Both are integrable models, which map to free fermions via the Jordan--Wigner transformation \cite{Giamarchi2003}. The entangling block in our architecture can be realized by such simple, native nearest-neighbor models, making the QSBM practical for analog quantum simulators. Despite the bare Hamiltonians being individually integrable, the multi-layer circuit, which alternates fixed unitary evolutions with independently trained, spatially inhomogeneous single-qubit rotations, effectively navigates the full scrambled regime, as confirmed by the convergence to Haar-typical KLD. This is directly verified in Appendix~\ref{sec:app_nonintegrable} by comparing both models against two non-integrable scramblers (mixed-field Ising and mixed-field XXZ). As shown in Fig.~\ref{fig:app_nonintegrable}, all three models collapse onto the same Haar-typical KLD limit at sufficient $\tau$ and $L$, confirming that Haar typicality, rather than the bare integrability of the Hamiltonian, is the primary physical resource.

\begin{figure}[t!]
    \centering
    \includegraphics[width=\columnwidth]{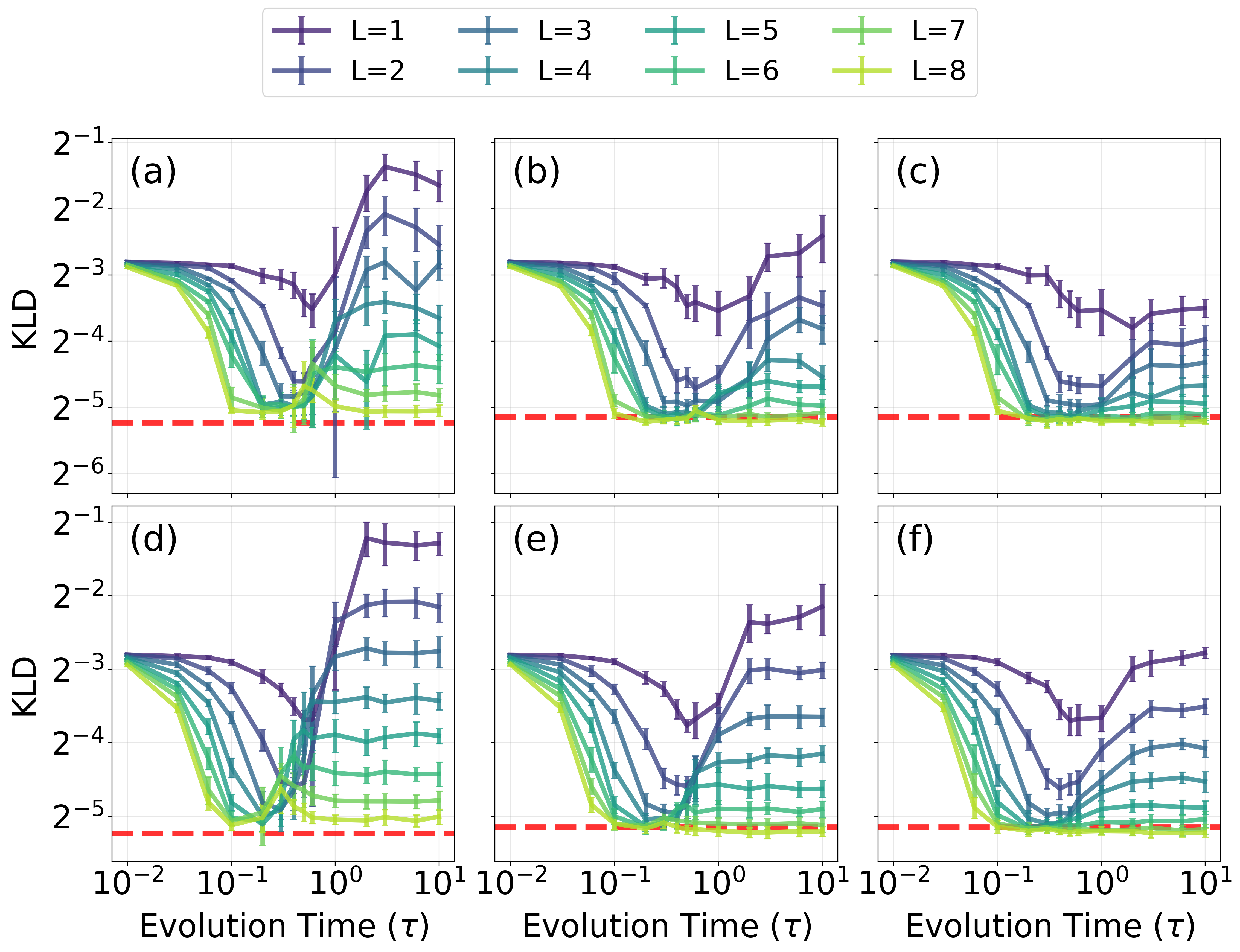}
    \caption{Analog Hamiltonian scramblers ($\NQ = 8$, $N_{\mathrm{shots}} = 5000$).
    Converged KLD vs.\ evolution time~$\tau$ for $L = 1$--$8$ layers (colored lines).   Columns: Ancilla qubits $\NA = 0, 1, 2$.
   Top row (a--c): Transverse-field Ising model [Eq.~\eqref{eq:H_tfim}]---the KLD drops at intermediate~$\tau \ge 0.1$ for $L>5$.
     Bottom row (d--f): XX model with transverse field [Eq.~\eqref{eq:H_xx}]---the KLD drops at $\tau\ge0.1$. Both Hamiltonians for $L \ge 5$ saturate at the Haar-scrambler KLD (red dashed) for $\NA = 1, 2$ ancilla qubits. The role of ancilla qubits is most visible for shallow circuits ($L<3$) for both Hamiltonians.}
    \label{fig:analog}
\end{figure}

In the following, we study QSBM performance as a function of evolution time $\tau$, for given number of layers $L$, and ancilla qubits $N_A$.
Figure~\ref{fig:analog} shows the converged KLD as a function of evolution time~$\tau$ for $L = 1$--$8$ layers and $\NA = 0, 1, 2$ ancillas, with the TFIM in the top row and the XX model in the bottom row.

At short evolution times ($\tau \ll 1$), $\Ua(\tau) \approx \hat{\mathbb{I}} - i\tau \Ho$ generates minimal entanglement, and the KLD saturates regardless of $L$ or Hamiltonian type. As $\tau$ increases, both models undergo a transition in which the KLD drops sharply around $\tau \sim 0.5$ (in units of the inverse coupling strength), coinciding with the half-chain entropy approaching the Page value.

The shallow circuits ($L\le4$) exhibit a non-monotonic dip-type dependence of the KLD on $\tau$: beyond the optimum, the maximally scrambled state has lost all mode structure of the target distribution, and few trainable single-qubit layers cannot restore the sharp inter-peak boundaries, causing the KLD to rise.
For deeper circuits, $L \ge 5$, the additional layers compensate for the excess entanglement and the non-monotonicity disappears. The quantum state is driven toward the maximally scrambled scenario, achieving performance consistent with the Haar limit (horizontal red dashed lines).

Consistently, allocating more qubits to the ancilla register (columns correspond to $N_A = 0, 1, 2$ from left to the right) enlarges the traced-out environment, so the scrambler-generated entanglement translates into a higher-rank mixed state of the sampler and a lower KLD over most of the $(\tau, L)$ grid; the reduction disappears at very short evolution times, where the state is barely entangled.

We note that the architecture confines trainable parameters to single-qubit rotations, while the entangling unitary is fixed and does not contribute gradients.

Gradient variance scaling is characterized in Appendix~\ref{sec:app_gradients} for $N = 4$--$14$, comparing the QSBM against a hardware-efficient ansatz (HEA) with fixed CNOT entanglers.
The variance decays as $O(2^{-N})$, consistent with the Lie algebraic prediction for global cost functions under $2$-design scrambling~\cite{ragone2024liealgebraic, cerezo2021cost, holmes2022connecting}.
The QSBM therefore does not evade barren plateaus, but at fixed parameter count matches the generative performance of standard parameterized QCBMs and the same gradient scaling as the equal-parameter HEA (Appendices~\ref{sec:app_gradients} and~\ref{sec:app_qcbm_comp}), while requiring no trainable entangling gates.

\section{Hamiltonian-driven generative modeling}\label{sec:2d}

We now extend the architecture by promoting the Hamiltonian parameters to trainable variables, replacing the pre- and post-rotation layers with trainable on-site fields $h_{x,i}^{(\ell)}$, $h_{y,i}^{(\ell)}$, $h_{z,i}^{(\ell)}$---which are equivalent to arbitrary single-qubit rotations at fixed evolution time---and adding a single trainable nearest-neighbor coupling $J_{xx}^{(\ell)}$ per layer.
Since the local fields can absorb any global rescaling, a trainable $J_{xx}^{(\ell)}$ at fixed~$\tau$ effectively controls the entangling strength per layer; with free local fields the rescaling is partly redundant, but in the discrete-layer ansatz the coupling and field parameters are not fully equivalent.

In contrast to the previous scramblers, where a single fixed unitary is reused in every layer, here each layer generates entanglement with an independently optimized coupling strength.
The full evolution thus corresponds to $L$ unitary evolutions driven by piecewise-constant Hamiltonians (computed via exact matrix exponentiation), where the initial product state is transformed into a final state whose Born-rule sampling in the computational basis reproduces the target distribution.

We test this variant on joint probability distributions $p(x, y)$ over two continuous variables.
The $\NQ = 10$ qubits (including $\NA = 2$ ancillas) are partitioned into two registers of $n_x = n_y = 4$ system qubits each; tracing out the ancillas yields a $2^{n_x} \times 2^{n_y}$ probability distribution.

\begin{figure}[t]
    \centering
    \includegraphics[width=\columnwidth]{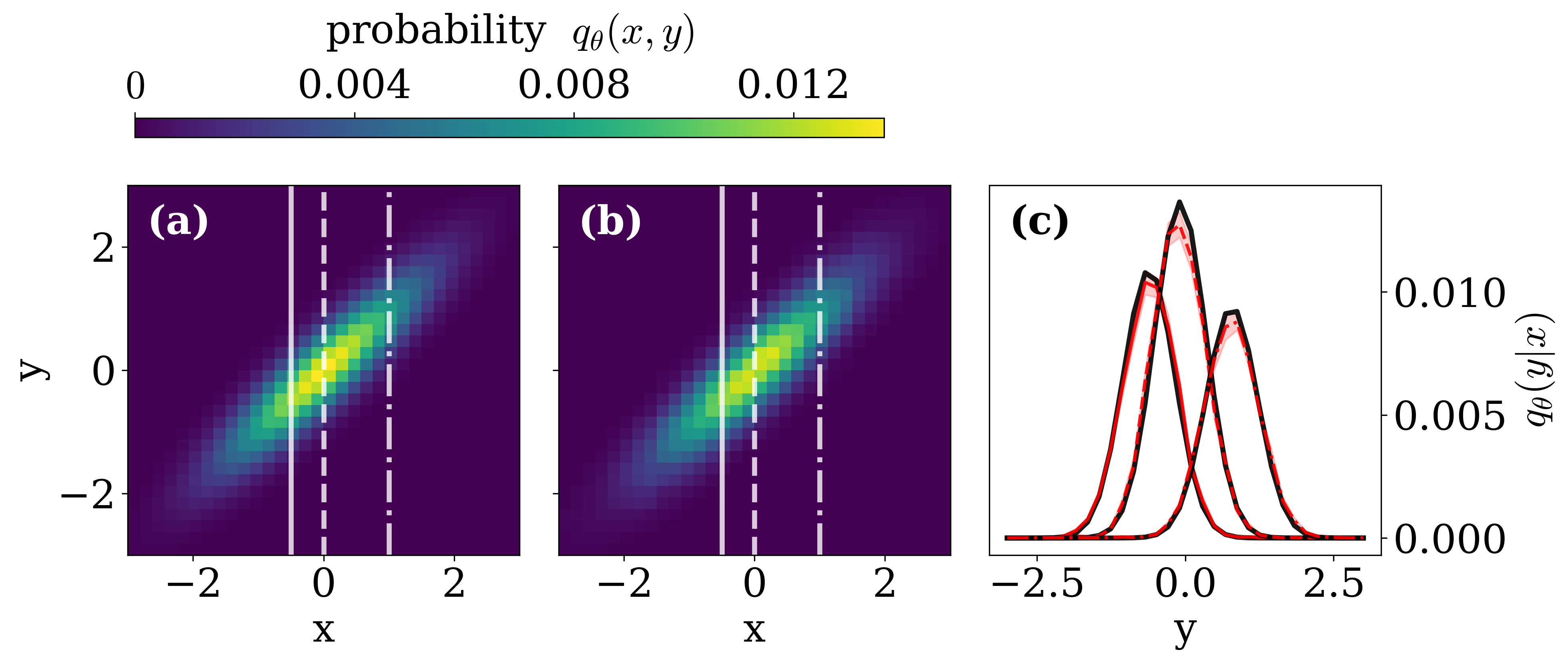}
    \caption{Trainable-Hamiltonian Born machine applied to the correlated 2D Gaussian of Eq.~\eqref{eq:2d_gauss} with $\rho = 0.9$ ($\NQ = 10$, $\NA = 2$, $n_x = n_y = 4$ qubits per register).
    The nearest-neighbor coupling $J_{xx}^{(\ell)}$ and local fields $h_{x,i}^{(\ell)}, h_{y,i}^{(\ell)}, h_{z,i}^{(\ell)}$ are independently optimized in each of $L = 10$ trainable layers ($\tau = 0.5$ per layer), yielding 310 trainable parameters.
    Averages over 20 realizations.
    {(a)}~Target distribution $p(x,y)$.
    {(b)}~Learned distribution $q_{\boldsymbol{\theta}}(x,y)$.
    {(c)}~Conditional slices $p(y\,|\,x)$ at three values of $x$ (dashed lines in panels a,b): target (black) vs.\ learned (red).}
    \label{fig:2d_gauss}
\end{figure}

We first consider the correlated bivariate Gaussian of Eq.~\eqref{eq:2d_gauss} with $\rho = 0.9$, which tests the model's ability to reproduce non-trivial inter-register correlations.
Figure~\ref{fig:2d_gauss} shows that the trainable-Hamiltonian QSBM faithfully captures both the elongated shape and the off-diagonal tilt of the target: panels~{(a)} and~{(b)} are visually indistinguishable, and the conditional slices in panel~{(c)} confirm quantitative agreement at three representative cuts through the distribution.

\begin{figure}[t]
    \centering
    \includegraphics[width=\columnwidth]{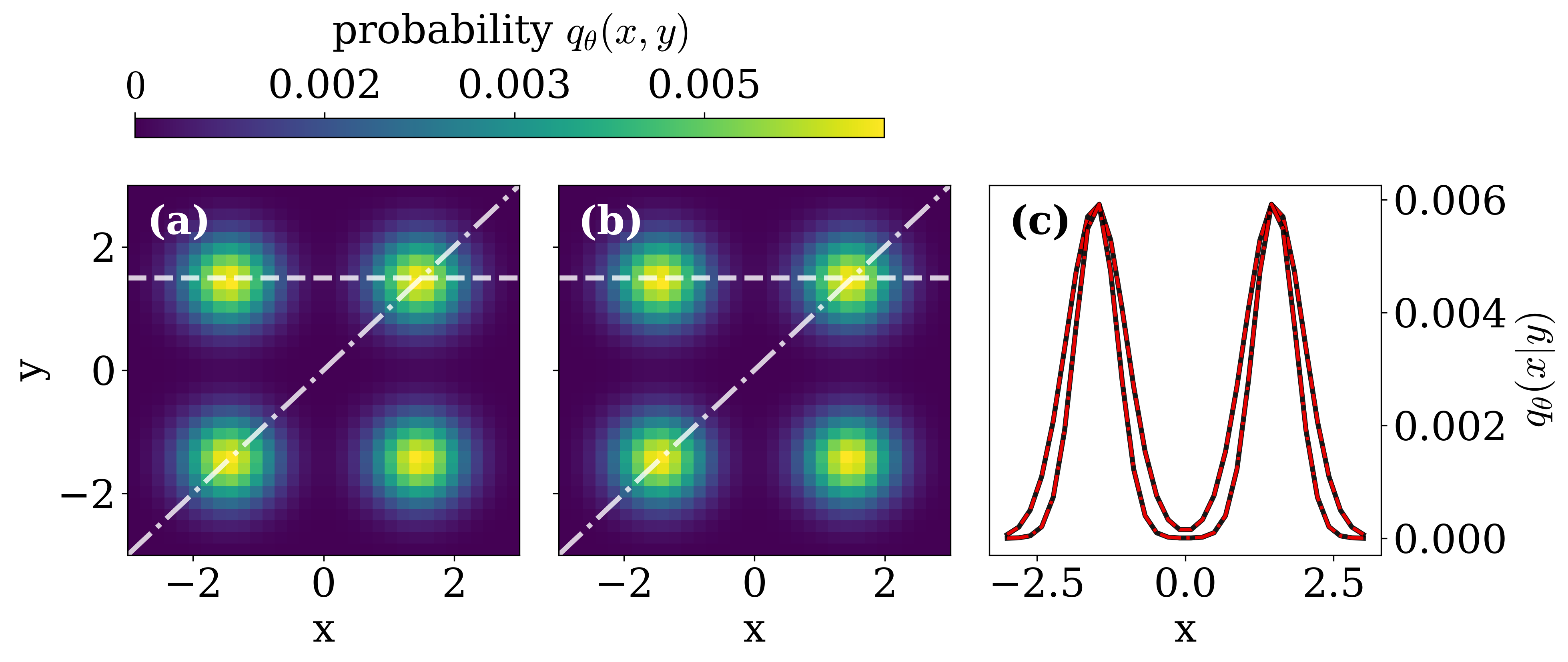}
    \caption{Trainable-Hamiltonian Born machine applied to the multimodal 2D target of Eq.~\eqref{eq:2d_mixture} ($\NQ = 10$, $\NA = 2$, $n_x = n_y = 4$ qubits per register).
    The nearest-neighbor coupling $J_{xx}^{(\ell)}$ and local fields $h_{x,i}^{(\ell)}, h_{y,i}^{(\ell)}, h_{z,i}^{(\ell)}$ are independently optimized in each of $L = 10$ trainable layers ($\tau = 0.5$ per layer), yielding 310 trainable parameters.
    Averages over 20 realizations.
    {(a)}~Target distribution.
    {(b)}~Learned distribution.
    {(c)}~Conditional slices through the peak locations: target (black) vs.\ learned (red).
    All four modes are captured without mode collapse.}
    \label{fig:2d_multimodal}
\end{figure}

The multimodal target of Eq.~\eqref{eq:2d_mixture} poses a more stringent test, as disconnected modes are notoriously prone to mode collapse in adversarial training~\cite{zoufal2019quantum}.
The QSBM captures all four modes with correct relative weights (Fig.~\ref{fig:2d_multimodal}), as confirmed by the conditional slices in panel~{(c)}.

\begin{figure}[t!]
    \centering
    \includegraphics[width=\columnwidth]{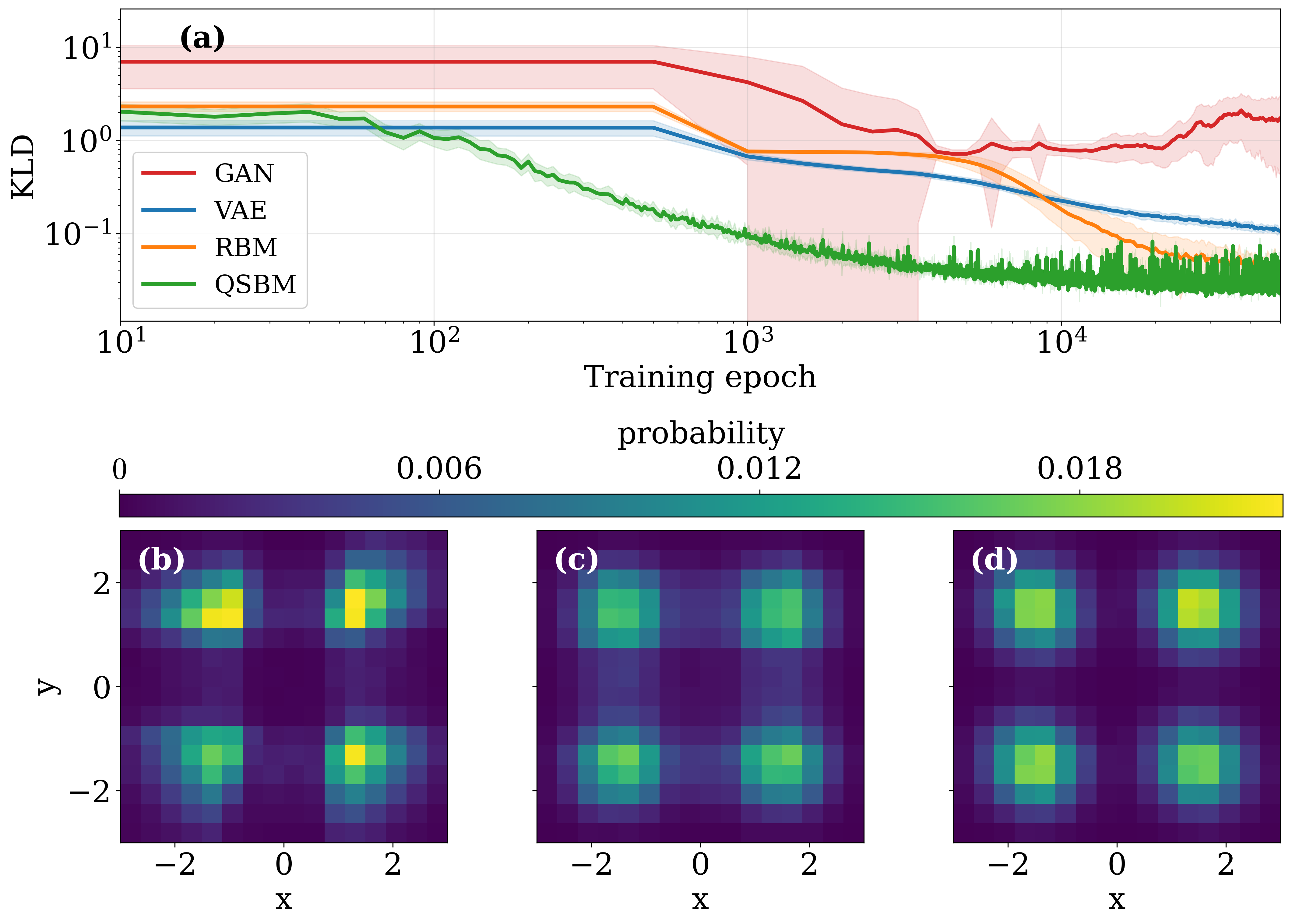}
    \caption{Top panel~{(a)}: KLD vs.\ training epoch for the trainable-Hamiltonian QSBM (310~params), a generative adversarial network (GAN), a variational autoencoder (VAE), and a restricted Boltzmann machine (RBM) on the multimodal target of Fig.~\ref{fig:2d_multimodal}, with matched parameter budgets ($\approx 310$) and $50\,000$ training epochs (mean $\pm 1\sigma$ over 20 realizations; log--log scale).
    Bottom panels: representative learned densities from the best-performing seed of each classical model.
    {(b)}~GAN (302~params; mean KLD: $1.74 \pm 1.30$).
    {(c)}~VAE (306~params; mean KLD: $0.11 \pm 0.01$).
    {(d)}~RBM (308~params; mean KLD: $0.045 \pm 0.008$).
    The QSBM (310~params) achieves a mean KLD of $0.030 \pm 0.005$.}
    \label{fig:classical}
\end{figure}

\subsection*{Benchmark against classical models}

We compare the trainable-Hamiltonian QSBM with three classical architectures at matched parameter budget ($\approx 310$): a generative adversarial network (GAN)~\cite{goodfellow2014generative}, a variational autoencoder (VAE)~\cite{kingma2014auto}, and a restricted Boltzmann machine (RBM)~\cite{hinton2012practical}.
This is not a quantum advantage claim; at this scale all models operate below the capacity needed for a fair generalization test~\cite{riofrio2024performance, hibatallah2024framework, bako2025fermionic}.
The comparison instead exposes qualitative failure modes: mode collapse in the GAN, over-smoothing in the VAE, and correct multimodal structure in the RBM.
A comparison against a fully parameterized QCBM is given in Appendix~\ref{sec:app_qcbm_comp}.

At this matched parameter budget, the training dynamics reveal differences among the QSBM and the considered classical models (Fig.~\ref{fig:classical}).
The top panel presents the mean KLD, averaged over
$20$ variational parameter initializations, vs.\ training epoch for each model.
The GAN exhibits large seed-to-seed variance and frequent mode collapse, reflecting the fragility of adversarial training at this parameter scale~\cite{goodfellow2014generative}.
The VAE converges reliably but produces axis-aligned, broadened peaks---artifacts of the Gaussian decoder and the KL regularization term that penalizes sharp latent encodings.
The RBM achieves the lowest KLD among the classical models. 

The QSBM distribution is normalized by construction via the Born rule, whereas classical models must enforce normalization explicitly (RBM partition function) or learn it implicitly (GAN, VAE evidence lower bound).
At this small parameter budget, the QSBM's advantage is structural: its entire parameter count is devoted to expressive single-qubit rotations around a fixed high-entanglement scrambler, while classical models at matched capacity must allocate parameters to encoder/decoder structure (VAE), adversarial balance (GAN), or partition-function normalization (RBM). Born-rule normalization is exact and requires no additional parameters. This advantage is expected to narrow as the parameter count grows and classical models gain sufficient capacity to overcome their structural overhead.

\section{Summary and conclusions}\label{sec:conclusions}

We have introduced a Quantum Scrambling Born Machine architecture that treats quantum scrambling---operationally defined here as the generation of Haar-typical bipartite entanglement---as a computational resource, where a fixed, non-trainable unitary generates all the entanglement while only single-qubit rotations are optimized. We considered an ideal Haar-random unitary and its two experimentally feasible approximations: finite-depth brickwork random circuits and analog evolution under nearest-neighbor spin-chain Hamiltonians.
For the benchmark distributions and system sizes studied ($\NQ \le 10$), once the entangling unitary approaches Haar-typical entanglement, the learned distribution closely matches the target with weak sensitivity to the scrambler's microscopic origin.
In the fixed-scrambler variant, the scrambler provides entanglement at zero trainable cost, so the entire parameter budget is devoted to expressivity, making the architecture particularly parameter-efficient and well suited to near-term platforms where the entangling operation---whether a native gate sequence or analog many-body dynamics---can be calibrated once and reused.
Tracing out ancilla qubits allows the model to explore mixed-state distributions, increasing its expressivity at fixed qubit budget. This observation suggests that noise inevitably present on physical hardware can be treated as an additional expressivity resource~\cite{sannia2024dissipation}, naturally introducing mixedness.

Finally, the trainable-Hamiltonian variant shows that Hamiltonian parameters can be optimized so that a piecewise-constant evolution, starting from a product state, generates a quantum state whose Born-rule measurement statistics reproduce a target classical probability distribution.
This casts many-body quantum dynamics as a programmable generative resource, complementing the inverse Hamiltonian learning problem---where unknown couplings are inferred from measurements---with a forward generative counterpart in which Hamiltonian parameters are optimized to reproduce a target classical distribution.

An interesting direction for future work is the implementation of mid-circuit measurements with classical feedforward.
Since the ancilla register is traced out at the end of each layer, this operation is mathematically equivalent to a deferred measurement without feedforward.
Replacing the partial trace with active mid-circuit measurements and conditional feedforward operations would enable adaptive state preparation within each layer.
This could provide a route to preparing non-unitary resources and exploiting measurement-driven shallow-circuit advantages for generative modeling~\cite{salvati2022introducing, cao2026measurement}.

\section*{Data availability}

The code that generates the data and figures presented in this article, together with the numerical data plotted in the figures, is openly available on GitHub~\cite{plodzien2026qsbmcode}.

\begin{acknowledgments}
We thank Grzegorz Rajchel-Mieldzioć and Anna Dawid for useful comments on the manuscript.
We acknowledge funds from MICIU/AEI/10.13039/501100011033/ FEDER, UE.
\end{acknowledgments}

\appendix

\section{Gradient Variance Scaling and Barren Plateaus}\label{sec:app_gradients}

\begin{figure}[t]
    \centering
    \includegraphics[width=\columnwidth]{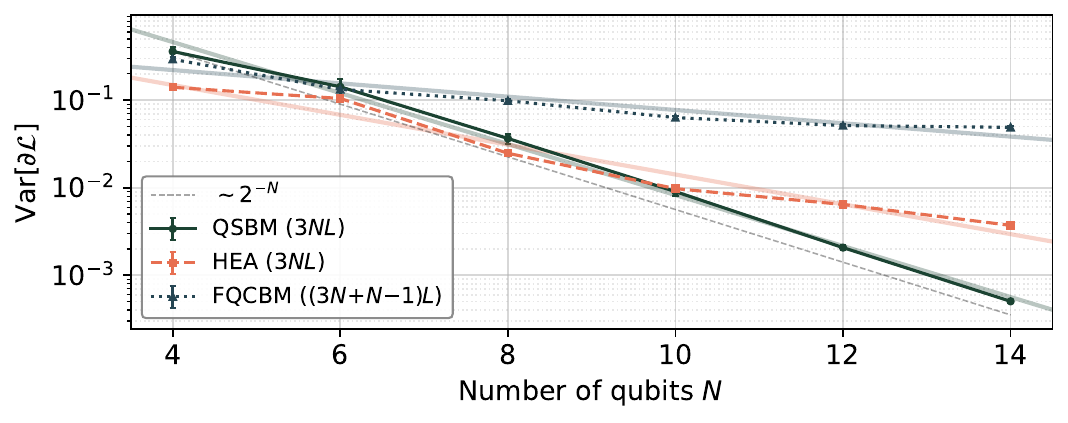}
    \caption{Gradient variance $\VarGrad$ vs.\ system size $N$ for the QSBM (RQC scrambler), the hardware-efficient ansatz (HEA, CNOT ladder), and the fully parameterized QCBM (FQCBM, $R_{ZZ}$ entangler) at $L = 4$.
    Markers with error bars: numerical data; lines: exponential fits; thin dashed: $2^{-N}$ reference.
    The QSBM and HEA track the $2^{-N}$ reference; the FQCBM decays more slowly.
    Averages over independent seeds.}
    \label{fig:app_gradients}
\end{figure}

We compute the gradient variance $\VarGrad$ by averaging over random parameter sets $\boldsymbol{\theta} \in [0, 2\pi)^{3NL}$ and, for the QSBM, scrambler realizations.
Three architectures are compared: the QSBM (brickwork RQC scrambler, depth $K=2N$), a hardware-efficient ansatz (HEA, fixed CNOT ladder, same parameter count $3NL$), and a fully parameterized QCBM (FQCBM, trainable $R_{ZZ}$ gates, $(3N+N-1)L$ parameters).
Fig.~\ref{fig:app_gradients} shows results for $N = 4, 6, 8, 10, 12, 14$ at $L = 4$.

The QSBM and HEA both decay as $O(2^{-N})$, consistent with the Lie algebraic bound $\VarGrad \le O(1/\dim\mathfrak{g})$~\cite{ragone2024liealgebraic}: a scrambler approaching a $2$-design generates the full DLA $\mathfrak{g} = \mathfrak{su}(2^N)$, giving $\VarGrad \le O(2^{-N})$ for global observables.
The FQCBM decays more slowly because its structured $R_{ZZ}$ entanglers produce a less Haar-typical ensemble at this depth.
The QSBM therefore does not evade barren plateaus, but achieves the same gradient scaling as the equal-parameter HEA without trainable entangling gates.

\begin{figure}[t]
    \centering
    \includegraphics[width=\columnwidth]{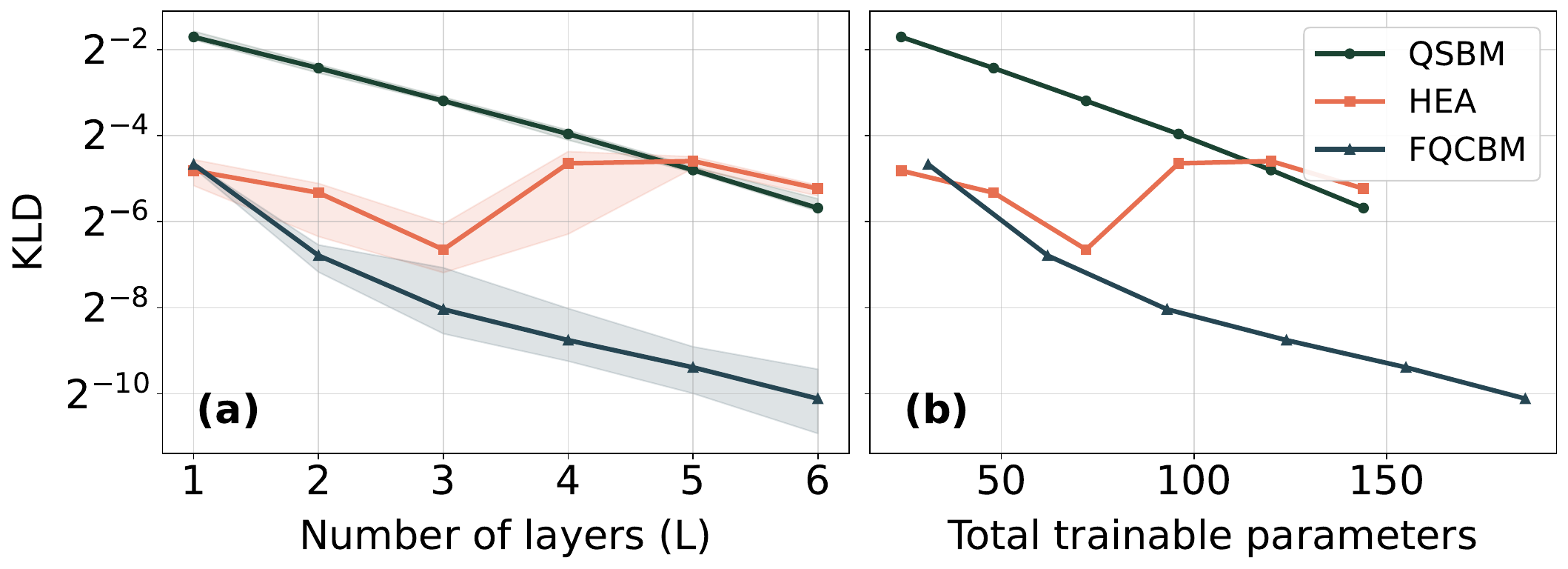}
    \caption{Comparison of the QSBM (brickwork RQC scrambler, depth $K = 2N$), the hardware-efficient ansatz (HEA, CNOT ladder), and the fully parameterized QCBM (FQCBM, trainable $R_{ZZ}$ gates) on the multimodal target [Eq.~\eqref{eq:multimodal}] at $N = 8$.
    {(a)}~Converged KLD vs.\ number of layers~$L$; lines show the median over 20 seeds and shaded bands the interquartile range.
    The HEA attains its best KLD at $L = 3$ and degrades for deeper circuits, whereas the QSBM decreases monotonically and matches or surpasses it for $L \ge 5$ at identical parameter count $3NL$.
    {(b)}~Median converged KLD vs.\ total trainable parameters.
    The FQCBM reaches the lowest KLD owing to its trainable two-qubit gates, at the cost of $(3N + N-1)L$ parameters.}
    \label{fig:app_qcbm_comp}
\end{figure}

\section{Comparison against Fully Parameterized QCBMs}\label{sec:app_qcbm_comp}

We isolate the effect of the fixed-scrambler architecture by comparing converged generative performance of the QSBM (brickwork RQC, depth $K=2N$), the hardware-efficient QCBM (HEA, CNOT ladder), and a fully parameterized QCBM (FQCBM, trainable $R_{ZZ}$ gates) on the multimodal target [Eq.~\eqref{eq:multimodal}] at $N = 8$.

As shown in Fig.~\ref{fig:app_qcbm_comp}(a), the HEA achieves its best KLD at $L = 3$ but degrades at deeper circuits, reflecting optimization difficulty of parameterized circuits with fixed entangling structure.
The QSBM decreases monotonically with $L$ and matches or surpasses the HEA at $L \ge 5$, at identical parameter count ($3NL$).
The FQCBM achieves the lowest KLD at large $L$ owing to trainable two-qubit gates, at the cost of $(3N + N-1)L$ parameters.
Fig.~\ref{fig:app_qcbm_comp}(b) plots KLD vs.\ parameter count. The QSBM and HEA share the same parameter count ($3NL$), with the QSBM surpassing the HEA at $L = 6$.
The QSBM reaches competitive performance with zero trainable entangling gates.

\section{Ancilla Scaling at Fixed Output Register}\label{sec:app_ancilla}

\begin{figure}[t]
    \centering
    \includegraphics[width=\columnwidth]{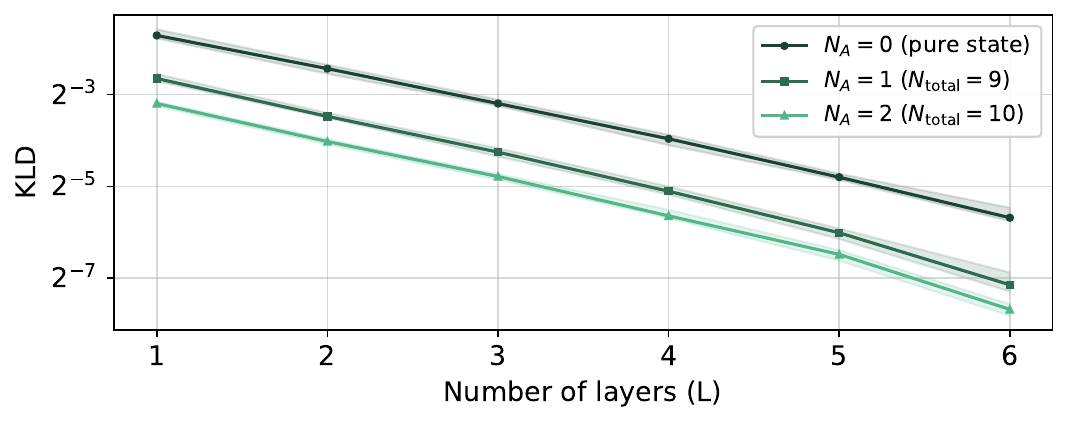}
    \caption{Ancilla effect at fixed output register size $n = 8$ for $N_A = 0, 1, 2$ (total system sizes $N_{\mathrm{total}} = 8, 9, 10$) using the Haar scrambler.
    Increasing $N_A$ systematically reduces the KLD at fixed target resolution, demonstrating that the mixed-state rank $2^{N_A}$ provides a genuine expressivity resource.
    Averages over 20 realizations.}
    \label{fig:app_ancilla}
\end{figure}

To resolve the influence of ancilla qubits on expressivity, we perform a controlled sweep at a fixed output register dimension $n = 8$ (discretized to $2^n = 256$ bins). 
We vary the number of ancilla qubits $N_A = 0, 1, 2$, which corresponds to total register sizes $N_{\mathrm{total}} = 8, 9, 10$. 
The scrambler operates on all $N_{\mathrm{total}}$ qubits, and the $N_A$ ancillas are traced out before measurement.
Fig.~\ref{fig:app_ancilla} shows the converged KLD vs. layer count $L$.

By holding the target distribution dimension fixed, we eliminate the dimensional reduction confound. 
The systematic decrease in KLD with larger $N_A$ shown in Fig.~\ref{fig:app_ancilla} confirms that the mixedness of the reduced density matrix $\hat{\rho}_{\mathrm{sys}}$, which has rank up to $2^{N_A}$, enlarges the representable class of distributions and acts as a genuine computational resource.

\section{Non-Integrable Hamiltonian Scramblers}\label{sec:app_nonintegrable}

In the main text, we considered the transverse-field Ising model, which we denote here as $\Ho_1 = \Ho_{\mathrm{TFIM}}$ [Eq.~\eqref{eq:H_tfim}], and the XX spin chain, both of which are Jordan--Wigner integrable in their bare forms. 
Here, we verify the insensitivity of the scrambled generative performance to the microscopics of the spin chain by introducing two non-integrable models:
(i) The mixed-field transverse-field Ising model (MFIM), denoted here as $\Ho_2 = \Ho_{\mathrm{MFIM}}$,
\begin{equation}\label{eq:H_mfim}
    \Ho_2 = -\sum_i \sz_i \sz_{i+1} - h_x\sum_i \sx_i - h_z\sum_i \sz_i,
\end{equation}
with $h_x = 1.0$ and $h_z = 0.7$; and
(ii) The mixed-field XXZ model, denoted here as $\Ho_3 = \Ho_{\mathrm{MFXXZ}}$,
\begin{equation}\label{eq:H_xxz_tilt}
    \Ho_3 = -\sum_i \left( \sx_i \sx_{i+1} + \sy_i \sy_{i+1} + \Delta \sz_i \sz_{i+1} \right) - h_x\sum_i \sx_i - h_z\sum_i \sz_i,
\end{equation}
with $\Delta = 0.6$, $h_x = 0.6$, and $h_z = 0.4$. 
Both models $\Ho_2$ and $\Ho_3$ are non-integrable and exhibit Wigner--Dyson level spacing statistics, characteristic of many-body quantum chaos.

\begin{figure}[t]
    \centering
    \includegraphics[width=\columnwidth]{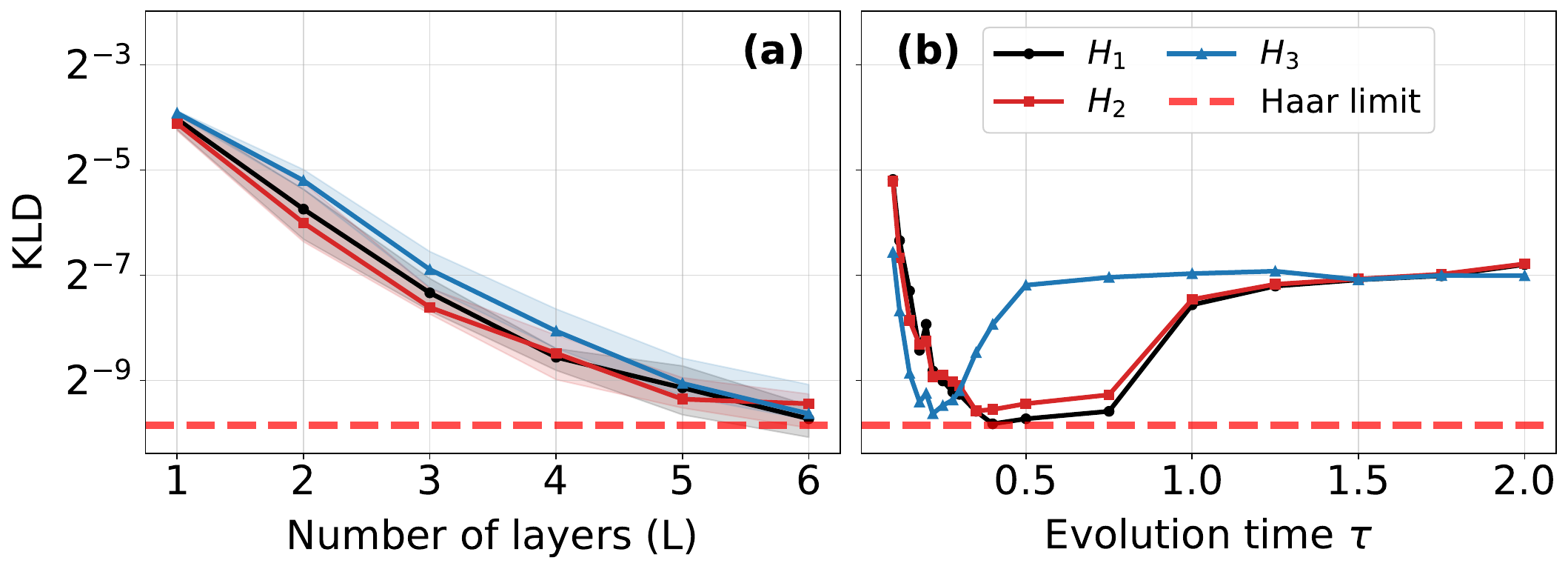}
    \caption{Comparison of integrable ($\Ho_1$, TFIM) and non-integrable ($\Ho_2$, MFIM; $\Ho_3$, mixed-field XXZ) scramblers at $N = 8$, $N_A = 1$.
    {(a)} Converged KLD vs.\ $L$ at representative $\tau$: $0.5$ for $\Ho_1$, $\Ho_2$ and $0.22$ for $\Ho_3$.
    {(b)} Converged KLD vs.\ $\tau$ at $L = 6$.}
    \label{fig:app_nonintegrable}
\end{figure}

All three analog Hamiltonians $\Ho_1$, $\Ho_2$, $\Ho_3$ yield qualitatively similar generative performance [Fig.~\ref{fig:app_nonintegrable}(b)].

\bibliography{references}

\end{document}